\documentclass[global,twocolumn]{svjour}
\journalname{Applied Physics B}
\usepackage[utf8]{inputenc}
\usepackage{graphicx}
\usepackage[leqno]{amsmath}
\usepackage{amsfonts}
\usepackage{amssymb}
\usepackage{bm}
\usepackage{hyperref}
\usepackage{color}
\usepackage{siunitx}
\usepackage{float}
\usepackage{flushend}

\begin{document}
\title{Multilayer ion trap with three-dimensional microwave circuitry for scalable quantum logic applications}

\author{H.~Hahn\inst{1,2}\thanks{HH and GZ contributed equally to this work.} \and G.~Zarantonello\inst{1,2}$^\star$ \and A.~Bautista-Salvador\inst{1,2,3} \and M.~Wahnschaffe\inst{1,2,3} \and M.~Kohnen\inst{1,2}\and J.~Schoebel\inst{4} \and P. O.~Schmidt\inst{1,2} \and C.~Ospelkaus \inst{1,2,3} \mail{christian.ospelkaus@iqo.uni-hannover.de} }
\institute{Physikalisch-Technische Bundesanstalt, Bundesallee 100, 38116 Braunschweig, Germany \and Institute of Quantum Optics, Leibniz Universität Hannover, Welfengarten 1, 30167 Hannover, Germany \and Laboratory for Nano- and Quantum Engineering, Leibniz Universität Hannover, Schneiderberg 39, 30167 Hannover, Germany \and Department for High-Frequency Technology, Technische Universität Braunschweig, Schleinitzstraße 22, 38106 Braunschweig, Germany}

\maketitle

\begin{abstract}
We present a multilayer surface-electrode ion trap with embedded 3D microwave circuitry for implementing entangling quantum logic gates. We discuss the electromagnetic full-wave simulation procedure that has led to the trap design and the characterization of the resulting microwave field-pattern using a single ion as a local field probe. The results agree with simulations within the uncertainty; compared to previous traps, this design reduces detrimental AC Zeeman shifts by three orders of magnitude. The design presented here can be viewed as an entangling gate component in a library for surface-electrode ion traps intended for quantum logic operations. 
\end{abstract}

\section{Introduction}
\label{sec:introduction}
Trapped ions are a promising platform to explore applications in quantum simulation and quantum computation~\cite{cirac_quantum_1995,blatt_entangled_2008,georgescu_quantum_2014,bermudez_assessing_2017}. Towards the ultimate goal of a large-scale universal quantum machine solving specific problems with a quantum speed-up~\cite{montanaro_quantum_2016}, milestones for first experimental implementations of quantum algorithms and quantum simulations have recently been achieved~\cite{debnath_demonstration_2016,monz_realization_2016,zhang_observation_2017,jurcevic_direct_2017}. However, in order to improve their practical use in substantial problems, the number of stored and manipulated qubits as well as the fidelity of operations have yet to be increased significantly, highlighting the remaining key ingredient of a scalable architecture~\cite{monroe_scaling_2013}. 

Surface-electrode traps~\cite{seidelin_microfabricated_2006}, where all electrodes are located in a plane, represent a suitable scalable platform as they allow to implement elements of the so-called 'quantum CCD' architecture~\cite{wineland_experimental_1998,kielpinski_architecture_2002}. Based on well-developed microfabrication techniques, surface traps unite an intrinsically scalable fabrication with a high degree of reproducibility. In such a planar geometry, the ion is trapped and influenced by the potentials that are applied to the metal electrodes located at the very top of the structure. Scaling to large arrays of ion traps~\cite{amini_toward_2010} will lead to complex electrode arrangements and thus make it indispensable to incorporate interconnections to lower layers embedded into the structure~\cite{craik_microwave_2014,guise_ball-grid_2015,maunz_high_2016}. Any such layers would be shielded from the ion by the top metal layer, but can be used to supply the electrodes in the top layer with control voltages. 

For universal qubit manipulation, single- and multi-qubit gates driven by microwave radiation~\cite{ospelkaus_trapped-ion_2008,mintert_ion-trap_2001} benefit from microfabricated traps as the control elements can be included as an integrated microwave conductor in the trap design. For qubits with transition frequencies in the RF or microwave regime, such as in $^{9}$Be$^{+}$, this enables direct access to the qubit transition and, for these atoms, thus avoids the necessity for a complex Raman laser system to carry out gates~\cite{gaebler_high-fidelity_2016,ballance_high-fidelity_2016}. In contrast to laser-based schemes, achievable fidelities of the microwave approaches~\cite{johanning_individual_2009,harty_high-fidelity_2016,ospelkaus_microwave_2011,khromova_designer_2012,weidt_trapped-ion_2016} show no fundamental limit due to photon scattering~\cite{ozeri_errors_2007} and are approaching the fault-tolerant regime for universal gate sequences~\cite{harty_high-fidelity_2014,harty_high-fidelity_2016}.  

In this paper we introduce a microfabricated radio-frequency ion trap with integrated 3D microwave circuitry. Besides two microwave conductors that can be used to drive single-qubit gates, the trap design features an embedded 3D microwave conductor with a meander-like shape that has been designed to carry out multi-qubit gates using microwave near-fields. The trap has been produced in a novel multilayer fabrication process enabling a scalable trap architecture based on microwave (near-field) quantum logic~\cite{bautista-salvador_multilayer_2019}.

In the following Sec.~\ref{sec:trap} we briefly introduce the fabrication process, trap design and experimental setup. Sec.~\ref{sec:microwave} treats the design of the 3D microwave conductor usable for multi-qubit gates in more detail and highlights its advantages in terms of scalability and field properties when compared to a corresponding single-layer trap. In addition, we give a description of the full-wave simulation model used for extracting its resulting magnetic field configuration. In Sec.~\ref{sec:exp} we compare these simulations to experimental results, including an S-parameter measurement and a Ramsey-type, single-ion experiment to map out the produced field configuration. Finally, in Sec.~\ref{sec:conclusions} we summarize and conclude our findings.  

\begin{figure*}[tb]
	\centering
	\includegraphics[width=\textwidth]{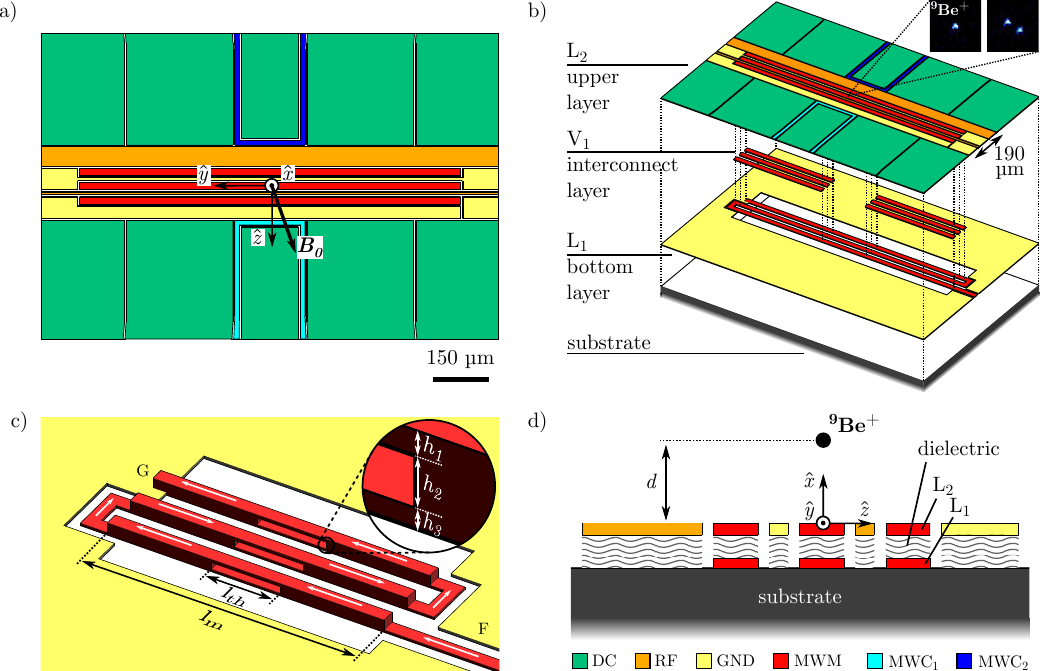}
	\caption{
		Multilayer trap layout a) Electrode configuration of the upper layer L$_2$ b) exploded-view of the three trap layers (dielectric material in L$_1$ and L$_2$ not shown). The inset gives the EMCCD camera signal for a single- and 2-ion crystal c) Schematic of the 3D meander-like microwave conductor MWM with distorted dimensions for clarity, showing its extension to the three fabrication layers L$_1$, V$_1$, and L$_2$ with respective thicknesses $h_1$, $h_2$ and $h_3$ (inset). White arrows indicate the direction of a hypothetic applied DC current as used in simulations for thermal effects (see appendix). Surrounding electrodes as well as the dielectric material have been excluded for illustration purposes d) cross section of the RF electrodes and the 3D microwave conductor MWM in the radial $xz$-plane at $y=0$. The dielectric material is used to isolate structures between the bottom and upper layer. For this trap configuration the ions are confined at $d\simeq\SI{35}{\micro\m}$ above the surface of L$_2$.
	}
	\label{fig:figure1}
\end{figure*}

\section{Trap design and fabrication}
\label{sec:trap}
The trap presented here is a multilayer extension of a surface-electrode ion trap with integrated microwave conductors~\cite{wahnschaffe_single-ion_2017} and consists of three individual fabrication layers called L$_1$, V$_1$ and L$_2$. Here, L$_1$ represents the bottom layer, L$_2$ the upper layer and V$_1$ the interconnect layer used to connect L$_1$ and L$_2$ where required (Fig.~\ref{fig:figure1}b). The fabrication process starts with a high-resistivity Si substrate coated with a $\SI{2}{\micro\m}$-thick insulating Si$_3$N$_4$ layer by plasma-enhanced chemical vapor deposition (PECVD). After thermally evaporating a \SI{10}{\nano\m}-thin layer of Ti and a \SI{50}{\nano\m}-thin layer of Au on top of the Si$_3$N$_4$ layer, the bottom layer L$_1$ is metallized by UV photolithography and a subsequent gold electroplating step. 

The metallization of the interconnect layer V$_1$ follows the same procedure as for L$_1$. After removing the Ti adhesion and Au seed layers with plasma etching, a dielectric film is spin coated on top of V$_1$ and L$_1$ (Fig. \ref{fig:figure1}d). Dielectric used is Polyimide (PI 2600 series, HD MicroSystems\texttrademark). A chemical-mechanical polishing (CMP) step is used to planarize the top dielectric surface. To ensure electrical contact between part of V$_1$ and L$_2$ we first perform a global etch-back process and stop close to the top part of V$_1$. Then we perform a local etch-back process on areas defined by photolithography on top of V$_1$

To form the upper layer L$_2$ we repeat the steps used for L$_1$; Au/Ti deposition, UV photolithography and gold electroplating. As a final step, via plasma etching, we remove the adhesion and seed layer between the electrodes which would otherwise short out all electrodes in L$_2$, and the dielectric underneath down to either L$_1$ or to the substrate.  For the trap described here, the resulting electrode thicknesses are $h_{1}=\SI{4.4}{\micro\m}$ for L$_1$, $h_{2}= \SI{9.5}{\micro\m}$ for V$_1$ and $h_{3}=\SI{5.2}{\micro\m}$ for L$_2$. The thicknesses values result from the optimization described in Sec.~\ref{sec:microwave}. The field required for electrical breakdown of the polymide is nominally $> 2\cdot 10^8\,\mathrm{V/m}$. For the voltage here reported no electrical breakdown is expected. For further details on the general fabrication method, see ref.~\cite{bautista-salvador_multilayer_2019}.

Figure~\ref{fig:figure1}a shows the upper layer L$_2$ which holds all electrodes relevant for ion trapping, given by two RF electrodes for radial confinement (both originating from a common feedline) and 10 DC electrodes for axial confinement. For this trap the resulting ion-to-electrode distance between the ions and L$_2$ is $d\simeq\SI{35}{\micro\m}$ (Fig.~\ref{fig:figure1}d). A radio frequency signal with frequency $\omega_{\mathrm{RF}}\simeq 2\pi\times 176.5\,\mathrm{MHz}$ and amplitude $V_{\mathrm{RF}}\simeq 100\,\mathrm{V}$ is applied to the RF electrode. The voltages applied to the DC electrodes range between $\pm26\,\mathrm{V}$. For the experiments perfomed in Sec.~\ref{sec:exp} this corresponds to secular trap frequencies in axial and radial direction of $(\omega_{\mathrm{ax}},\omega_{\mathrm{LF}},\omega_{\mathrm{HF}})\simeq 2 \pi \times (4.12,5.6,9.33)\,\mathrm{MHz}$ for a single $^{9}$Be$^{+}$ ion. Based on calculations using the gapless plane approximation~\cite{wesenberg_electrostatics_2008} the high-frequency (HF) radial mode forms an angle of $-5.9^\circ$ with respect to the $x$-axis and the intrinsic trap depth is $\mathrm{10\,meV}$. 

\begin{figure}[tb]
	\centering
	\includegraphics[width=0.9\columnwidth]{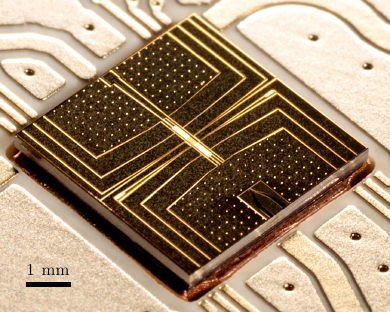}
	\caption{
		Photograph of the presented multilayer trap surrounded by a custom printed circuit board before wirebonding. The feedline of MWM lies completely in L$_1$ and starts with a tapered bonding pad centered at the bottom right side. In the trap center the conductor is partly extended to L$_2$. Multiple vias in V$_1$ that interconnect ground electrodes in L$_1$ and L$_2$ are seen as bright dots.
	}
	\label{fig:figuretrap}
\end{figure}

In contrast to the electrodes needed for ion trapping, the 3D microwave conductor labeled MWM in Fig.~\ref{fig:figure1} is extended to all three fabrication layers (L$_1$, V$_1$, L$_2$), enabling a complex microwave conductor design, which is discussed in more detail in Sec.~\ref{sec:microwave}. It is designed to produce an oscillating magnetic near-field gradient at the ion position suitable to drive motional sidebands (the key ingredient for multi-qubit gates) on the first-order field-independent qubit transition $\left|F=2,\,m_F=+1\right>\equiv\left|\uparrow\right> \leftrightarrow  \left|F=1,\,m_F=+1\right>\equiv\left|\downarrow\right>$ of the electronic ground state $^{2}S_{1/2}$ with $\omega_0\simeq 2 \pi \times 1082.55\,\mathrm{MHz}$ at $|\bm{B_{0}}|\simeq 22.3\,\mathrm{mT}$ (see Fig. 2 in~\cite{wahnschaffe_single-ion_2017} for a level scheme). Here, $F$  refers to the total angular momentum $\bm{F}$ and $m_F$ is the quantum number of its projection on $\bm{B_{0}}$. The transition  $\left|2,0 \right> \leftrightarrow \left|1,0\right>$ has a frequency of  $\omega_1\simeq 2 \pi \times 1397.56\,\mathrm{MHz}$, this transition will be relevant in Sec.~\ref{sec:exp}. The two additional microwave conductors labeled MWC$_1$ and MWC$_2$ produce an oscillating magnetic field amplitude at the ion position and can each be used to induce carrier transitions in the $^{2}S_{1/2}$ hyperfine manifold with $\Delta m_{F} \in \{0,\pm1\}$. The grounded electrodes in the upper and bottom layer are connected via multiple interconnects in V$_1$, see Fig.~\ref{fig:figuretrap}. The number of interconnects is kept high to increase the conductivity between the layers.

The trap is mounted and wirebonded to a custom printed circuit board (PCB) placed in a room temperature vacuum system with pressure around $1 \times 10^{-11}\,\mathrm{mbar}$. The PCB features $50\,\mathrm{\Omega}$ coplanar microwave waveguides and RC filters ($f_{\mathrm{c}}\simeq 194\, \mathrm{kHz}$) for each DC trace. The Polyimide has a maximum long time baking temperature of $250\,\mathrm{C}^\circ$; more stringent constraints to the used maximum baking temperature in our setup come from in-vacuum microwave components. The trap is loaded by an ablation loading scheme~\cite{leibrandt_laser_2007} utilizing nanosecond pulses at 1064 nm to create a neutral $^{9}$Be ablation plume above the trap center and a 235 nm cw laser beam for subsequent photoionization~\cite{lo_all-solid-state_2013}. Doppler cooling is performed on the cycling transition $\left|^{2}S_{1/2},2,+2\right>$ $\leftrightarrow$ $\left|^{2}P_{3/2},m_J=+3/2,m_I=+3/2\right>$ using  $\sigma^{+}$ polarized light at $\simeq$ 313 nm~\cite{wilson_750-mw_2011} propagating parallel to $\bm{B_{0}}$, forming an angle of $30^\circ$ with the $z$-axis.

\section{Three-dimensional microwave conductor}
\label{sec:microwave}

\subsection{Microwave conductor design}
\label{subsec:design}
In the microwave near-field approach the Rabi rate of motional sideband transitions is proportional to the oscillating magnetic field gradient $B^\prime$ multiplied by the ion wavepacket size $x_{\mathrm{wp}}=\sqrt{\hbar/(2m\omega_\mathrm{m})}$ where $m$ is the ion mass and $\omega_\mathrm{m}$ the motional mode frequency ($x_{\mathrm{wp}}\approx 8\,\mathrm{nm}$ for $\omega_\mathrm{m}=\omega_{\mathrm{HF}}=2 \pi \times 9.33\,\mathrm{MHz}$). Whereas a high magnetic field gradient increases the motional coupling speed, any residual magnetic field amplitude $B$ produced at the ion position along the gradient can degrade the spin-motion coupling fidelity by off-resonant carrier excitations and/or uncompensated AC Zeeman shifts~\cite{harty_high-fidelity_2016}. Consequently, the main design criterion of the microwave conductor MWM introduced in Fig.~\ref{fig:figure1} is to produce a magnetic field configuration which maximizes $B^\prime$ at the ion position while keeping $B$ as small as possible. 

As shown in previous work, a single microwave conductor with a meander-like shape~\cite{carsjens_surface-electrode_2014} can fulfill these field requirements by producing an oscillating magnetic quadrupole whose central field minimum is overlapped with the ion position at the RF null position. The minimum of the magnetic quadrupole is defined as the point where the amplitude of the magnetic field oscillation is minimized. Compared to a three-conductor design~\cite{ospelkaus_microwave_2011}, the single meander-shaped conductor eliminates position fluctuations of the microwave magnetic field minimum due to phase and amplitude instabilities between the independently driven conductors of~\cite{ospelkaus_microwave_2011}. The meander shaped MWM conductor design for the present trap is sketched in Fig.~\ref{fig:figure1}c (not to scale).

As can be seen, the feedline coming from the position labeled `F' as well as the turning parts of the meander structure itself are completely in the bottom layer L$_1$, while MWM extends to all trap layers (L$_1$, V$_1$, L$_2$) along the three segments of length $l_{\mathrm{m}}$. The turning points are moved away from the segments and placed underneath the ground electrodes of L$_2$. This will shield the connection segments from the ion(s).  At the center of each segment, there is a pocket of length $l_{\mathrm{th}}=\SI{200}{\micro\m}$ and thickness $h_2$ which is filled with dielectric material during the fabrication process (see Fig.~\ref{fig:figure1}d for a cross-section through the pocket center labeled `X' in Fig.~\ref{fig:figure1}c). At its end, the conductor is terminated to a ground patch in L$_1$ at position labeled `G'. The main advantages of the multilayer fabrication and the 3D meander-like conductor over equivalent single-layer trap designs are summarized in the following.

First, we demonstrate the possibility to bring in signals, such as the microwave signal for MWM, in the bottom layer L$_1$ and to connect them to trap electrodes controlling the ion only where needed. This avoids having to put these signal paths in the upper layer in L$_2$, where they would interfere with other trap electrodes, and is highly desirable for scalability. The same approach can also be applied to DC and RF electrodes.

Second, the fact that the different segments of the MWM conductor are only connected in the bottom layer L$_1$ allows us to independently choose the length of the RF electrode and the length $l_{\mathrm{m}}$ of the MWM segments. To achieve the desired overlap of the RF and microwave field minima, at least one RF electrode needs to be placed between MWM segments. For comparison, in the single-layer design of ~\cite{wahnschaffe_single-ion_2017}, this constrained the length of the RF electrodes to be less than the length of the MWM segments. It is, however, desirable to extend the RF electrodes much further along the axial direction in order to be able to transport ions between the entangling gate trap module presented here and other trap zones which would then be part of a surface-electrode ion trap array implementing the `quantum CCD architecture'~\cite{wineland_experimental_1998,kielpinski_architecture_2002}.

Lastly, the implementation of the pocket inside the meander segments allows, in this configuration, to reduce the residual magnetic field $B$ at the ion position by roughly one order of magnitude while keeping the gradient unchanged~\footnote{For a simulated input power of 1$\,W$ we obtain $B=\SI{7}{\micro T}$ without a thermal gap and $B=\SI{0.8}{\micro T}$ with a thermal gap of length $l_{\mathrm{th}}=\SI{200}{\micro\m}$; the gradient $B^\prime=28\,\mathrm{T/m}$ remains unchanged in both cases.}, thus dramatically improving the field properties of MWM. No simple physical picture has been found for this decrease. In the discussed design this is possible only with a 3-dimensional architecture of MWM, no other comparable magnetic field suppression was found by varying parameters which affect the arrangement of the electrodes in the trap plane. From numerical simulations we have investigated the influence of the pocket length $l_{\mathrm{th}}$ on both the thermal load and the residual field $B$. For the latter we have not found further suppression for values higher than $l_{\mathrm{th}}=\SI{200}{\micro\m}$. We perform resistive heating simulation to determine the thermal behaviour of MWM. This is of importance because the pocket is filled by Polyimide which is a worse thermal conductor than gold. We simulate the effect of a DC current of  $1\,$A and measure the temperature at the position marked by X in Fig.~\ref{fig:figure1}c. As can be intuitively understood the heating is smaller for smaller sizes of $l_{\mathrm{th}}$. A detailed discussion of the thermal load in the trap caused by operating the MWM conductor can be found in the appendix.

\subsection{Simulation model}
\label{subsec:simulation}
Before starting any simulation of the MWM conductor design as discussed in the next paragraph, a decision about the desired ion-to-electrode distance $d$ has to be made. The following points are particularly relevant here: on the one hand, the achievable gradient at the ion position scales as $d^{-2}$, consequently allowing high multi-qubit gate speeds for small distances. On the other hand, the motional heating rate $\dot{\bar{n}}$ should be expected to scale as $d^{-4}$~\cite{turchette_heating_2000}, making the heating rate a significant error source during multi-qubit gates for small values of $d$. In our case, we consider an ion-to-electrode distance of $d\simeq\SI{35}{\micro m}$ to be a good compromise. This distance was selected to enhance the gradient as much as possible while still being at a distance which was known to work for other surface-electrode trap experiments.

\begin{figure}[tb]
	\centering
	\includegraphics[width=0.8\columnwidth]{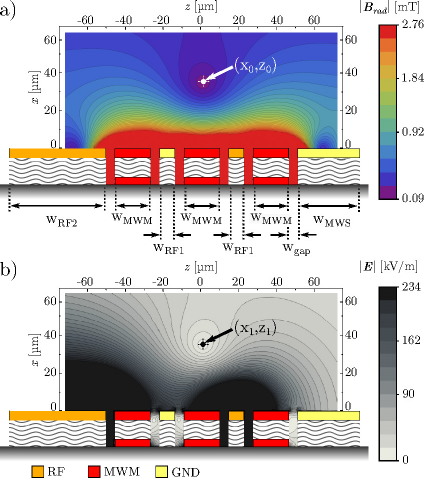}
	\caption{
		a) Simulated  magnetic field pattern in the radial $x$-$z$ plane produced by the microwave conductor MWM for 1 W of excitation. The dimensions indicated below are swept in simulations in order to overlap the magnetic field minimum at $(x_{\mathrm{0}},z_{\mathrm{0}})$ with the ion position. b) Simulated electric field pattern produced by the RF electrodes for 1 W of excitation. The ions are trapped in the RF null position at $(x_{\mathrm{1}},z_{\mathrm{1}})$. For a perfect overlap of the two indicated positions, the gradient $B^\prime$ is maximized and the residual field $B$ minimized at the ion(s) position.
	}
	\label{fig:figure3}
\end{figure}

The simulation model we employ to predict the resulting field configuration of a specific MWM conductor design around the qubit transition frequency ($\simeq1\,\mathrm{GHz}$) is based on full-wave finite-element simulations using Ansys HFSS 17.2 software and addresses high-frequency effects like the skin-depth, eddy currents or couplings between neighbouring conductors. Its overall purpose is to find a MWM conductor design maximizing $B^\prime$ while minimizing $B$ at the ion position by overlapping the magnetic field minimum at $(x_{0},\,z_{0})_{y=0}$ with the RF null position at $(x_{1},\,z_{1})_{y=0}$ as shown in Fig.~\ref{fig:figure3}. 

All simulations are performed with a radiation boundary condition, assuming a gold conductivity at room temperature of $4.1\cdot10^7\,\mathrm{Siemens/m}$ for all electrodes and conductors and include wirebonds as well as a section of the surrounding PCB. The nominal loss tangent of the polyimide is $0.002$ at $1\,\mathrm{kHz}$ which is also the parameter used in the simulation, it is assumed constant at all frequencies to simplify the simulation. No strong change in results has been observed for the variation of this parameter. The input of MWM is simulated with a $50\,\mathrm{\Omega}$ waveport while lumped ports are used for the inputs of the RF electrode as well as MWC$_1$ and MWC$_2$. The RF lumped port has a complex impedance of $100+1000i\,\mathrm{\Omega}$, the carrier ones have an impedance of $50\,\mathrm{\Omega}$ as each is connected to a $50\,\mathrm{\Omega}$ wave\-guide. For the RF port we have assumed a mostly inductive behavior reflected in the high reactance with some losses. A typical simulation workflow includes parametric sweeps of all the parameters given in Fig.~\ref{fig:figure1}c and Fig.~\ref{fig:figure3}a and can be structured in the following three steps.

In the first step, we only aim to find a starting geometry that coarsely fulfills the aforementioned field properties, namely an overlap of the magnetic field minimum and the RF null position about $\SI{35}{\micro m}$ above the trap. Here multiple electrode shapes, which differ from the one in Fig.~\ref{fig:figure1}c, can be tested. The location of the magnetic field minimum is found to depend most strongly on the spacing between the three MWM segments of length $l_{\mathrm{m}}$, their width $w_{\mathrm{MWM}}$, the width $w_{\mathrm{MWS}}$ of the left ground electrode and the width $w_{\mathrm{RF2}}$ of RF2 (see Fig.~\ref{fig:figure3}a). Analogously, the RF null position depends on the spacing between RF1 and RF2 as well as their individual widths $w_{\mathrm{RF1}}$ and $w_{\mathrm{RF2}}$. For each iteration the RF null position is calculated using the gapless plane approximation~\cite{wesenberg_electrostatics_2008} while the behaviour of the magnetic field minimum is inferred from multiple coarse parameter sweeps using the finite-element software. For this specific configuration the spacing between the three MWM segments is kept symmetric, therefore the distance between the segments is equal to two times the gap size $w_{\mathrm{gap}}$ plus $w_{\mathrm{RF_1}}$. No additional residual field cancellations effect have been observed in breaking this symmetry.

Once a coarse geometry has been found, the second step is to perform fine sweeps over all parameters with priority to further minimizing the residual field $B$ (even at the cost of an increased mismatch between the RF null and the magnetic field minimum of up to 1-2 $\SI{}{\micro\m}$).

In the last step the likely present mismatch is addressed by changing only parameters with a known impact on the magnetic field minimum or the RF null position. For instance, $w_{\mathrm{RF2}}$ and the width of the large ground electrode $w_{\mathrm{MWS}}$ can be used to minimize a minimum mismatch along the the $z$-axis, while the upper layer's thickness $h_3$ mostly affects the magnetic field minimum along the $x$-axis. Due to increased sensitivity requirements, we determine the RF null position in this last step by using finite element simulations with the correct frequency applied to the RF electrode. In our simulation model we consider a final mismach between $(x_{0},\,z_{0})_{y=0}$ and $(x_{1},\,z_{1})_{y=0}$ of smaller than $100\,\mathrm{nm}$ acceptable.

\subsection{Couplings between conductors}
\label{subsec:coupling}
\begin{figure}[tb]
	\centering
	\includegraphics[width=\columnwidth]{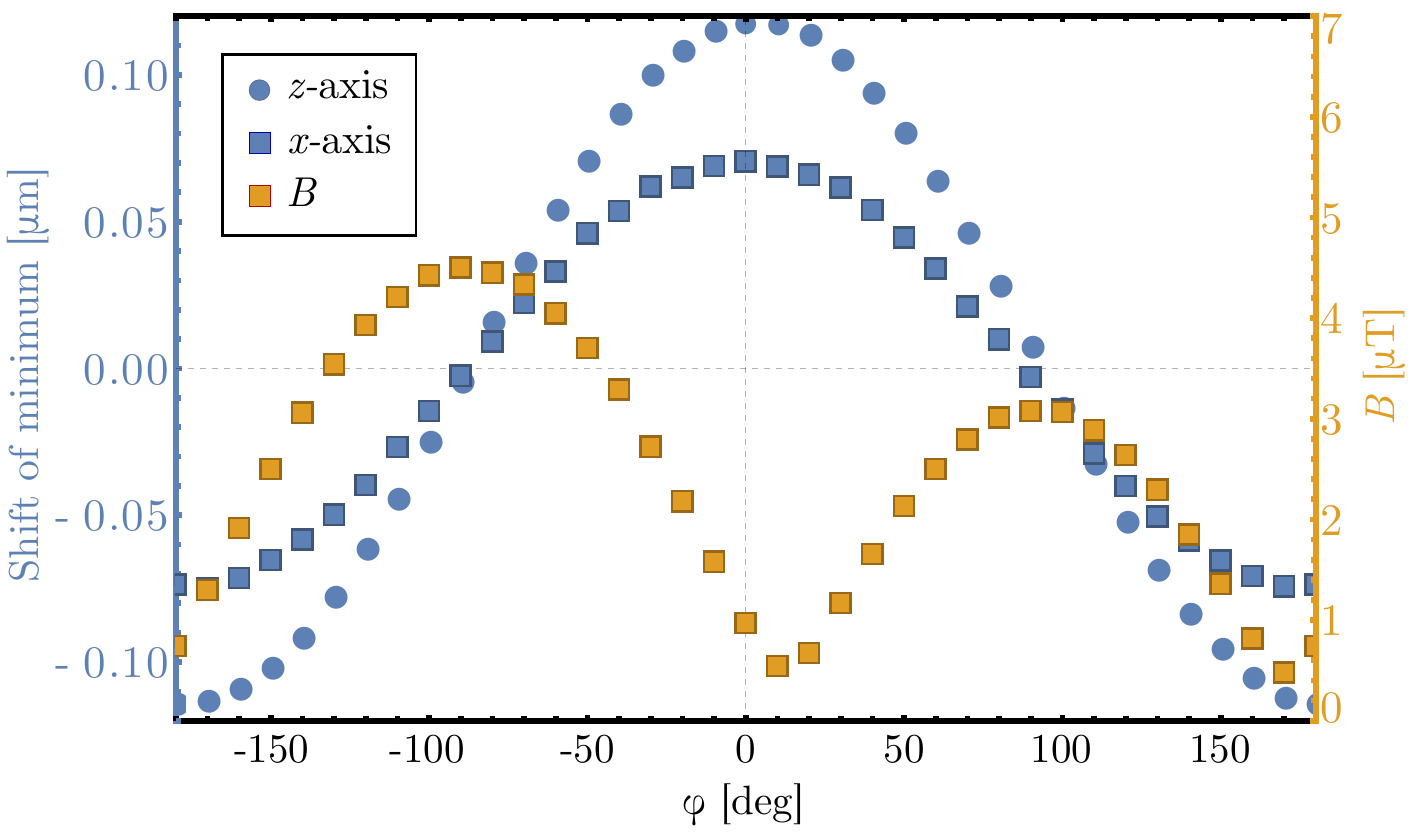}
	\caption{
		Effect of the -27.8 dB coupling between the MWM and MWC$_2$ conductors, at $1\,$W input power in MWM, for different phases of the backreflection on the transmission line attached to MWC$_2$. Blue markers: effect on the magnetic field minimum position of the quadrupole produced by MWM. Orange markers: effect on the total residual magnetic field $B$ due to additional magnetic fields produced by backreflections in MWC$_2$ 
	}
		\label{fig:figure7}
\end{figure}

The presence of multiple conductors constitutes a complex problem which the FEM simulation model can not fully account for due to its limited size. The issue arises because of the inductive and resistive coupling between electrodes and the possibility of back reflections of the induced currents in those electrodes. It affects mostly the field configuration created by MWM since any back reflected current can change the magnetic near-field configuration and hence can disturb the overlap with the ion's position. Since RF and DC are floating electrodes, the effect is negligible. On the other hand, the coupling to the microwave conductors MWC$_1$ and MWC$_2$, see Fig.~\ref{fig:figure1}, can cause notable changes of the minimum in MWM magnetic field that depends on coupling parameters and the transmission line of the other conductor~\cite{wahnschaffe_single-ion_2017}. This can be explained in the following way. The signal in MWM produces currents in MWC$_1$ and MWC$_2$ which propagate along the transmission line of each electrode. Some fraction of these induced currents is  backreflected at every impedance variation or imperfection in the transmission line, each with a certain phase depending on the transmission line properties. Once the sum of reflected currents has traveled back to the original electrode, it generates magnetic fields which have to be added to the magnetic field initially produced by MWM, causing an effective shift of the magnetic quadrupole minimum position. The simulation model used here includes only a section of the PCB, so any reflection which does not take place there or on the trap is not accounted for. To evaluate the impact of these reflections we consider a single perfect backreflection from the transmission line which reflects all the current sent to it. The effect is implemented by applying a signal to the corresponding lumped port exactly equal to the power coupled into it by MWM. This constitutes a worst-case scenario since the backreflected current can not be higher than the one originally coupled into an electrode. To understand the possible variation range it is necessary to analyze what happens for different phases of the backreflection. The blue markers (circles for $z$-axis, squares for $x$-axis) in Fig.~\ref{fig:figure7} show the position of the magnetic field minimum as a function of the phase of the backreflected current in MWC$_2$ for a coupling of $-27.8\,\mathrm{dB}$ at an input power of $1\,$W in MWM (the effect of MWC$_1$ is neglected here for clarity, but would just be added to the field of MWC$_2$). The change of the residual magnetic field $B$ at the magnetic minimum position is shown by the orange markers in Fig.~\ref{fig:figure7}. Even though for some phases the additional field further suppresses the overall residual field, for most phases the total value of $B$ increases.

In the previous trap design mentioned above, the coupling between MWM and the neighbouring microwave conductor, given by the S$_{21}$ parameter, was simulated to be $-9\,\mathrm{dB}$. The resulting position changes of the magnetic field minimum were up to $\SI{3}{\micro\m}$ and made it necessary to minimize the effect by engineering the transmission line appropriately. Since a reason of the strong coupling was found to be the immediate proximity of both involved conductors (compare Fig. 1 in~\cite{wahnschaffe_single-ion_2017}), we took special care in the conductor layout of MWC$_1$ and MWC$_2$ in the present multilayer trap. Here, the two microwave conductors are routed around the central DC electrode on each side which minimizes the length of closest proximity and allows a higher spatial separation of MWM and MWC$_1$ by introducing an additional ground electrode of width $w_{\mathrm{MWS}}$ (see Fig.~\ref{fig:figure3}a).
Following this approach, the coupling from MWM to MWC$_1$ and MWC$_2$ could be reduced by roughly $19\,\mathrm{dB}$ compared to the previous trap design to simulated values of $-27.9\,\mathrm{dB}$ and $-27.8\,\mathrm{dB}$, respectively.  
MWC$_1$ and MWC$_2$ are connected through wirebonds to the filterboard and are grounded at the other end via interconnects in V$_1$, which provide connection to a ground electrode in L$_1$. With the described configuration the coupling could be reduced to a level where its effect can be neglected in the fields overlap predictions, so no further increase of isolation was investigated.  

\section{Experimental data}
\label{sec:exp}
To measure the magnetic near-field pattern produced by MWM and compare it to our simulation model, we measure the induced AC Zeeman energy shift on suitable transitions in the atomic hyperfine structure for different ion positions in the radial plane. The exact procedure is detailed in~\cite{wahnschaffe_single-ion_2017} and based on previous work~\cite{Warring_2012}. The basic idea is to off-resonantly excite MWM between two $\frac{\pi}{2}$ pulses of a standard Ramsey experiment and to measure the induced AC Zeeman shift on the transitions  $\left|F=2,\,m_F=+1 \right> \leftrightarrow \left|1,+1\right>$ (qubit) and $\left|2,0 \right> \leftrightarrow \left|1,0\right>$ of the electronic ground state $^{2}S_{1/2}$ by fitting the observed phase accumulations due to the induced transition frequency change. The MWM conductor is excited with  a frequency of  $\omega\simeq 2 \pi \times 1092.55\,\mathrm{MHz}$. While the shift on the qubit transition is mainly induced by the $\pi$-component of the excitation field, the shift on the other transition is also induced by its $\sigma$-components. Moving the ion along the $x$- and $z$-direction by applying DC potentials allows to map the AC Zeeman energy shifts in the radial plane and to infer the resulting magnetic field distribution produced by MWM.

\begin{figure}[tb]
	\centering
	\includegraphics[width=\columnwidth]{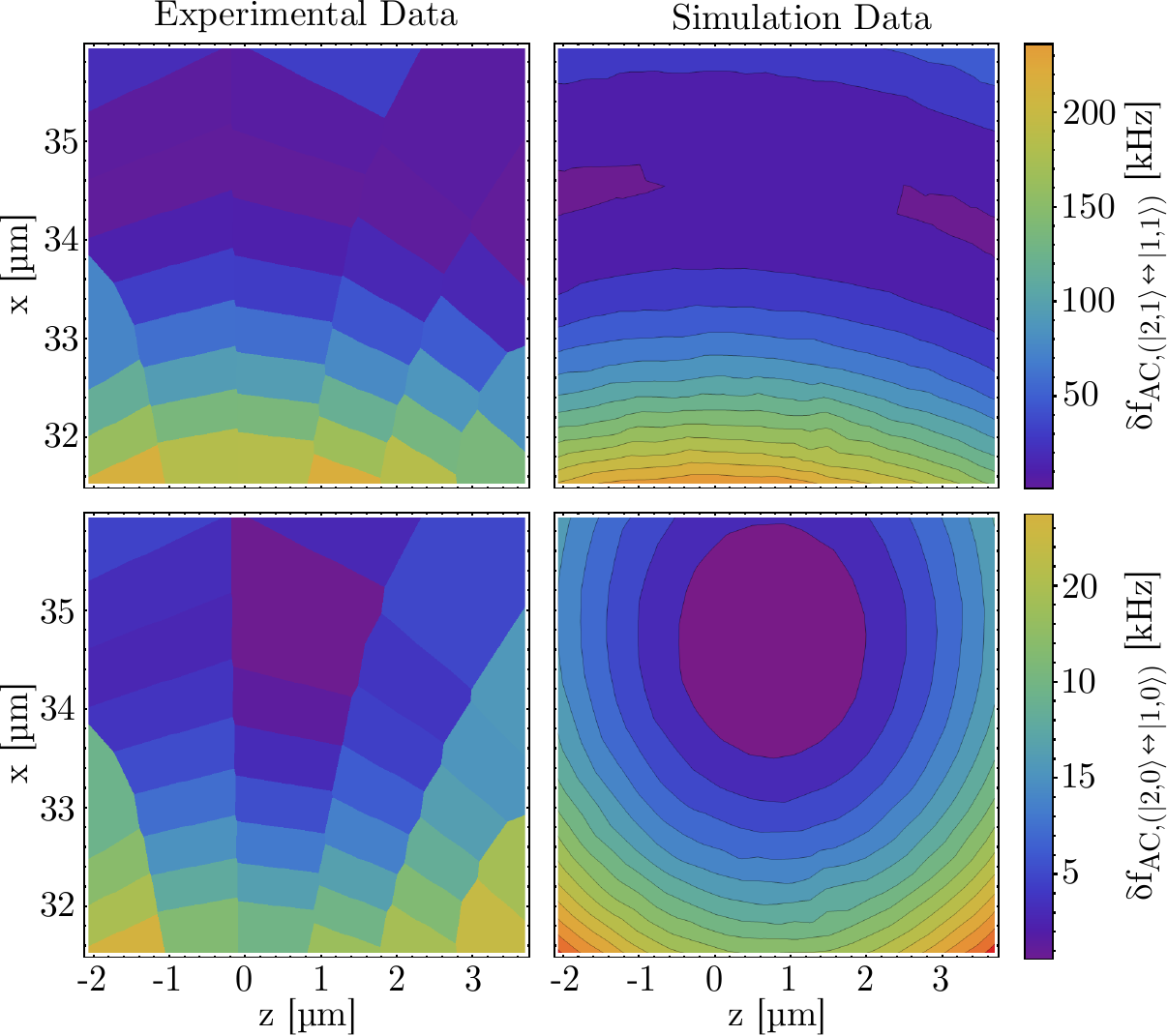}
	\caption{Comparison of experimentally measured (left column) and simulated (right column) absolute AC Zeeman shifts induced by the resulting field of the microwave conductor MWM. The upper row shows the data for the qubit transition $\left|2,+1 \right> \leftrightarrow \left|1,+1\right>$ while the bottom row shows the one for the $\left|2,0 \right> \leftrightarrow \left|1,0\right>$ transition. The position of the experimental data is relative to the position of the RF null which, in the coordinate system of Fig.~\ref{fig:figure1}, appears nominally at  $(x_1,\,z_1)=(0.75,\,34.55)\SI{}{\micro m}$. See main text for details.
	}
	\label{fig:figure6}
\end{figure}

Fig.~\ref{fig:figure6} shows the absolute values of measured AC Zeeman shifts on the qubit transition $\left|2,+1 \right> \leftrightarrow \left|1,+1\right>$ (upper row) and the $\left|2,0 \right> \leftrightarrow \left|1,0\right>$ transition (lower row) as a function of $x$ and $z$ around the calculated RF null position at $(x_1,\,z_1)=(0.75,\,34.55)\SI{}{\micro m}$. While the left column displays experimentally measured data, the right column shows AC Zeeman shifts calculated from the magnetic fields obtained by the simulation model introduced in Sec.~\ref{subsec:simulation}. The AC Zeeman shifts of the experimental data and the magnetic field of the simulations are each fitted in a single least squares fit using the 2D quadrupole model of Eq. (2) in~\cite{wahnschaffe_single-ion_2017} with $B$, $B^\prime$, $\alpha$, $\beta$, $\psi$, $x_\mathrm{0}$ and $z_\mathrm{0}$ as fit parameters. Here, $B$ is the residual field in the minimum, $B^\prime$ the magnetic field gradient, $\alpha$ the rotation angle of the residual field, $\beta$ the rotation angle of the gradient in the quadrupole, $\psi$ the relative angle between real and imaginary part and ($x_0$,$z_0$) the magnetic field minimum position. Since the residual field in the minimum of the experimental data was too low to be fitted accurately, $B$ and $\alpha$ were manually set to zero in the fit procedure. In consequence, only an upper bound for $B$ can be given. A comparison of the fit parameters determined in the experiment and by simulations can be found in Table~\ref{tab:table1}. The upper bound for $B$ is obtained by introducing it again in the fitted model as a parameter and by calculating the $B$ required to obtain the lowest measured AC Zeeman shift on the $\left|2,0 \right> \leftrightarrow \left|1,0\right>$ transition given by $551 \, \mathrm{Hz}$. For the calculation we assume that the measured shift is located at the absolute minimum. The value for $\alpha$ is set to $0^\circ$ since it minimizes the coupling, consequentely giving us an upper boundary for the residual field $B$ of $\sim\SI{34}{\micro T}$. We note that a smaller boundary could possibly be achieved by increasing the spatial resolution of the measurement.

\begin{table}[tb]
		\begin{tabular}{lll}
			\textbf{Parameter [units]}  			& \textbf{Simulation}            & \textbf{Experiment}\\ \hline
			$B \,[\SI{}{\micro T}]$     	& $1.47$   				& $\lesssim34$\\
			$B^\prime \,[\mathrm{T/m}]$	& $54.8$   				& $54.8(1.2)$\\
			$\alpha \,[^\circ]$   		& $40.8$          		& $-$\\
			$\beta \,[^\circ]$  	  	& $87.3$         		& $86.8(1.7)$\\
			$\psi \,[^\circ]$     		& $0.1$          		& $1.5(7.6)$\\
			$x_0 \,[\SI{}{\micro\m}]$   	& $34.72$  				& $34.62(0.05)$\\
			$z_0 \,[\SI{}{\micro\m}]$     & $0.73$			  	& $0.6(0.7)$\\ \hline
		\end{tabular}
	\caption{Comparison of the 2D quadrupole parameters determined by simulations and by experimental measurements using a single $^{9}$Be$^{+}$ ion (Fig.~\ref{fig:figure6})}
	\label{tab:table1} 
\end{table}

To induce higher frequency shifts, the data in the lower row was taken and simulated at a nominally $3\,\textrm{dB}$ higher power level than the data set in the upper row. For reference, the data in the upper row in Fig.~\ref{fig:figure6} corresponds to $B^\prime\approx54.8\,\textrm{T/m}$ obtained with $1.9\, \mathrm{W}$ of input power to the system.

\begin{figure*}[tb]
	\centering
	\includegraphics[width=\textwidth]{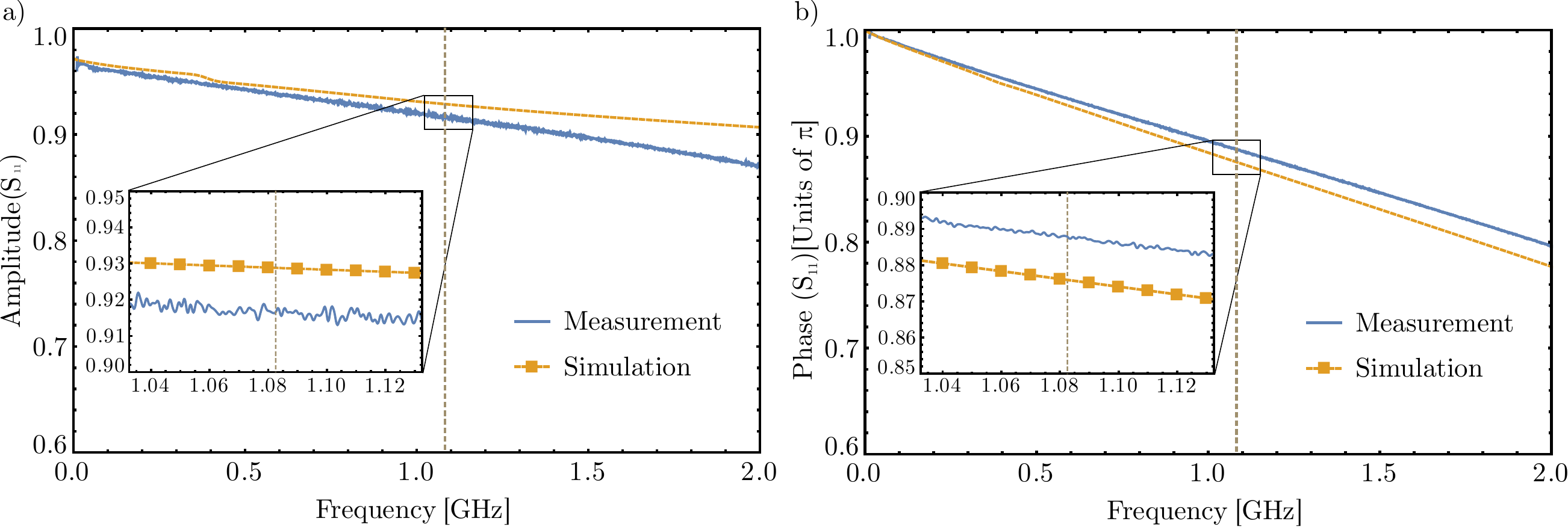}
	\caption{
		Comparison of MWM's measured and simulated S$_{11}$ parameter in amplitude (Fig.~a)) and phase (Fig.~b)) as a function of frequency. Measurement in blue simulations in dashed orange. The frequency of the qubit transition is highlighted by the dashed grey line. At that frequency we measure a reflected amplitude of $0.916$ while from simulation we expect $0.927$. The inset show a more detailed plot around the qubit frequency.  
	}
	\label{fig:figure8}
\end{figure*}

Complementary information can be obtained by analyzing the structure in terms of a microwave circuit. Our full-wave numerical simulations provide the S parameters of the structure for all implemented ports. Using a wafer prober, we can compare this data to actual measurements on fabricated devices (see Fig.~\ref{fig:figure8}). Within the frequency regime of interest (around 1 GHz), simulations and experimental data are in good agreement. The deviations for higher frequencies are not relevant for our experiments and can be explained by subtleties of the wafer prober measurements at higher frequencies, where the exact method of contacting the prober to the sample becomes more relevant and slightly deviates from the scenario employed in the simulations. Note that here we re-run the field simulations with all electrodes floating except for MWM and without the filterboard in order to mimic the conditions of the measurement. 

\section{Summary and conclusion} 
\label{sec:conclusions}
We have presented a multilayer surface-electrode ion trap based on a novel fabrication technique~\cite{bautista-salvador_multilayer_2019}. The trap design features two microwave conductors (MWC$_{1}$, MWC$_{2}$) to produce a high field amplitude for driving single-qubit gates and one 3D meander-like microwave conductor (MWM) to produce an oscillating near-field gradient with minimal residual field for driving multi-qubit gates around 1 GHz. 
Besides a general trap characterization, we give detailed information about the employed full-wave simulation model and the design criteria of the microwave conductors in order to produce the desired field configuration. The resulting field pattern of MWM was experimentally measured in a Ramsey-type single-ion experiment and subsequently analyzed in a least square model fit. The extracted 2D quadupole properties of the experimental data were found to be in good agreement with the expected values from our simulation model close to the ion position. 
The 3D microwave conductor extends to all three fabrication layers and allows an advanced version of a meander-shaped microwave conductor design when compared to a corresponding single-layer design. In the demonstrated multilayer trap, the 3D conductor enables an improved field pattern for multi-qubit gates by further suppressing the residual magnetic field $B$ at the ion's position. For comparison, the lowest residual field at a gradient of $54.8\,\textrm{T/m}$ in this trap was determined to be $\sim\SI{34}{\micro T}$ while the residual field in a previous single layer design~\cite{wahnschaffe_single-ion_2017} was calculated to be at least a factor of 14 larger ($\SI{447}{\micro T}$) when scaled to the same gradient.
The resulting AC Zeeman shift can cause significant error contributions during multi-qubit gates~\cite{harty_high-fidelity_2016} and depends on the residual field's magnitude and polarization. Analyzing the data sets of both traps, we found the absolute lowest measured AC Zeeman shift for the qubit transition in the multilayer trap ($697\,\mathrm{Hz}$) to be almost 3 orders of magnitude smaller than in the previous single-layer trap ($505\,\mathrm{kHz}$) when scaled to the same gradient.

Assuming the residual field of $\lesssim\SI{34}{\micro T}$ to be perfectly $\pi$-polarized we expect at the quadrupole minimum an on-resonance Carrier pi-time longer than $\SI{1.5}{\micro s}$ for the qubit transition. Based on the measured values of the gradient and the HF radial mode~\ref{sec:trap}, we calculate the minimum time for an entangling gate operation based on the M{\o}lmer-S{\o}rensen interaction~\cite{sorensen_quantum_1999,molmer_multiparticle_1999} to be $\approx \SI{120}{\micro s}$. Future work will focus on achieving control over the motion at the single-quantum level, measure heating rates and employ such devices for the implementation of entangling gates and effective spin-spin interactions.

\begin{acknowledgement}
We acknowledge support by the PTB cleanroom facility team and funding from PTB, QUEST, LUH, NTH (project number 2.2.11) and DFG through CRC 1227 DQ-\textit{mat}, project A01.   
\end{acknowledgement}

\appendix
\section{Joule heating}

\label{subsec:dissipation}
When microwave currents are applied to the trap through a conductor with finite conductivity, electric energy is converted into heat through resistive losses. Excessive resistive heating in the trap might lead to a degeneration of its performance or, in extreme cases, to an irreversible damage of the conductor. This is especially critical for an ion trap with multiple layers as discussed here, since current-carrying conductors defined around the geometric trap center have relatively narrow lateral dimensions, i.e. poor thermal contact to the substrate.

To include thermal effects caused by applying short microwave pulses with high power to the MWM conductor imitating real experimental conditions, one would need to numerically solve a complex system, including the whole chip geometry and inductive heating caused on neighbouring electrodes. However, it is also possible to gain insight into thermal effects in the trap design by using a simplified model in which a constant DC current is applied to MWM. As already introduced in Sec.~\ref{subsec:design}, we make use of a pocket in the three segments of MWM with length $l_\mathrm{m}$ in order to further decrease the residual magnetic field $B$ at the ion's position. Since we are interested in a compromise between a sufficiently low heat load and low $B$, the simplified DC power model still reveals the qualitative behaviour of the system.

In the experiment we apply we apply $1.9\,$W to the MWM structure corresponding to c.a. $160\,$mW of power not reflected by the trap given the measured S-parameter, see caption of Fig.~\ref{fig:figure8}. To better understand the limits of thermal dissipation we perform simulations on a DC current in a similar structure. We use a current of $1\,$A which corresponds to a dissipated power of $200\,$mW since the nominal geometry considered in the model, $l_{\mathrm{th}}=\SI{200}{\micro\m}$, has a resistance of $0.2\,\mathrm{\Omega}$.

\begin{figure}[ht]
	\centering
	\vspace{1cm}
	\includegraphics[width=0.9\columnwidth]{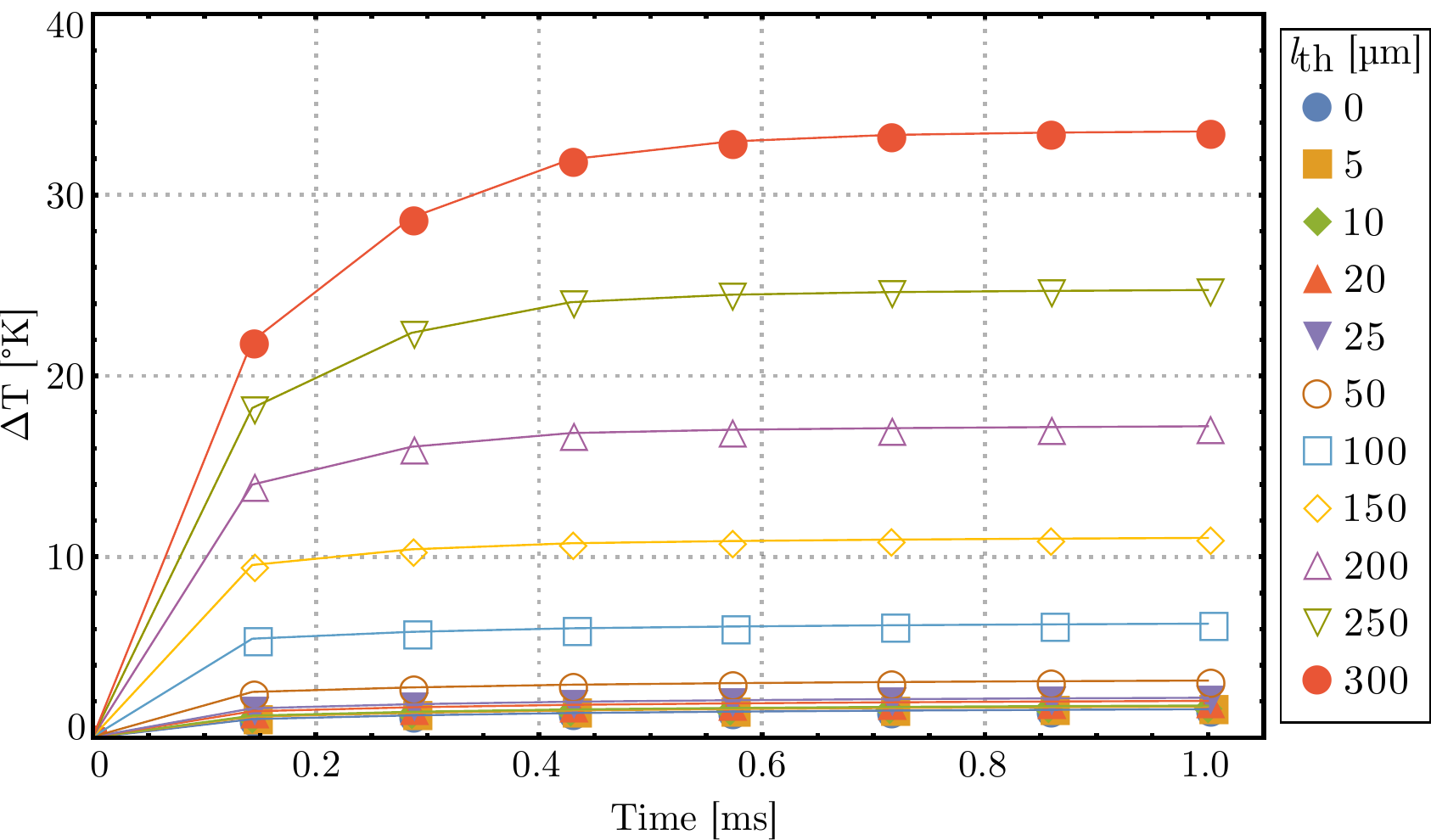}
	\caption{Results of the simplified finite-element model in which a constant current of $1\,$A is applied to MWM. The plots show the temperature increase in the center of MWM as a function of excitation time. Different colors indicate a different pocket length $l_{\mathrm{th}}$. 
	}
	\label{fig:figure5}
\end{figure}

The numerical simulations are performed using the AC/DC and heat transfer modules of COMSOL 4.3, assuming a DC current of \SI{1}{\ampere} applied to MWM, supported on an Si$_3$N$_4$/Si substrate of $\SI{1.5}{\mm}\times\SI{1.5}{\mm}$. We perform a parametric sweep of $l_{\mathrm{th}}$ along the $y$-axis. As a result of the fabrication process, the pockets are filled with a dielectric material, which we included in the model assuming a thermal conductivity of \SI{0.15}{\watt\per\meter\kelvin}. The electrical and thermal properties of Au (conductor material) and Si$_3$N$_4$ (wafer material) are taken from the built-in materials data library. While the room temperature ($T_0 = \SI{293.15}{\kelvin} $) reference is defined to be at the backside of the chip trap, the temperature of MWM is monitored on the surface of L$_2$ at the geometric trap center marked as `X' in Fig.~\ref{fig:figure1}c. At this position the heat load is maximal because each segment of MWM vertically splits into two parts of thickness $h_1$ and $h_3$, respectively, separated by the pocket of thickness $h_2$ in the interconnect layer V$_1$ filled with the dielectric material (see inset Fig.~\ref{fig:figure1}c). Naturally, at this point the heat will be poorly transported along the $x$-axis perperdicular to the trap. 

\begin{figure}[htb]
	\centering
	\vspace{1cm}
	\includegraphics[width=0.8\columnwidth]{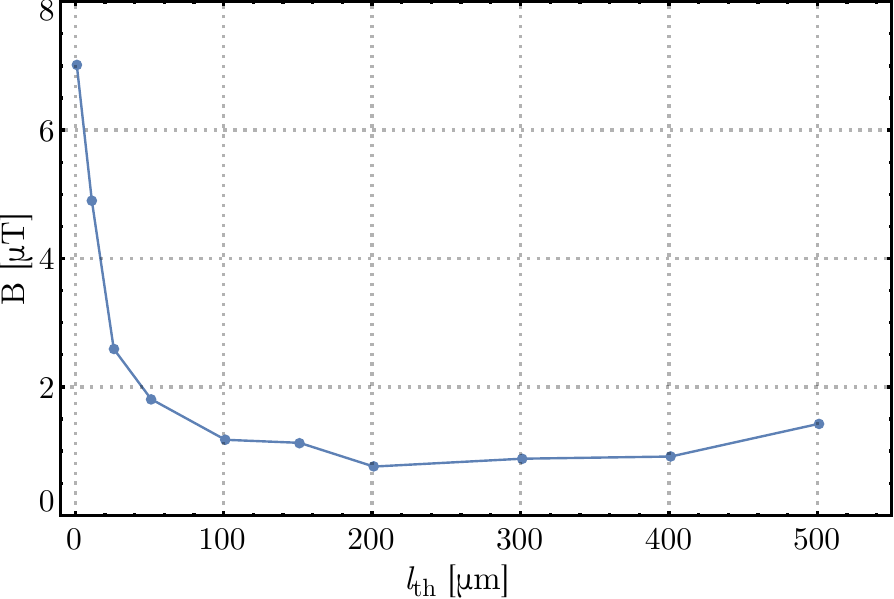}
	\caption{Residual magnetic field $B$ for different pocket lengths $l_{\mathrm{th}}$ in the segments of the MWM conductor. At about $l_{\mathrm{th}}=\SI{200}{\micro m}$ there is no further decrease in the residual field. The values were obtained by performing FEM simulations using our simulation model discussed in Sec.~\ref{subsec:simulation}. 
	}
	\label{fig:figure9}
\end{figure}

Fig.~\ref{fig:figure5} shows the resulting temperature change with respect to room temperature for different values of $l_{\mathrm{th}}$ as a function of time. For $l_{\mathrm{th}}=\SI{200}{\micro\meter}$, the system reaches a steady state after \SI{0.6}{\milli\second}, inscreasing its temperature by \SI{17}{\kelvin}. As shorter values for $l_{\mathrm{th}}$ decrease the temperature change, but increase $B$ at the ion position (see Sec.~\ref{subsec:design}), we found $l_{\mathrm{th}}=\SI{200}{\micro\m}$ to be the best compromise as illustrated in Fig.~\ref{fig:figure9}.

\end{document}